\documentclass[aps,prl,twocolumn,epsfig,floats]{revtex4}
\usepackage{graphicx,epsf,natbib,amsmath,latexsym,amssymb}
\def\be{\begin{equation}}
\def\ee{\end{equation}}
\def\bea{\begin{eqnarray}}
\def\eea{\end{eqnarray}}

\begin{document}


\title{Adler-Bell-Jackiw anomaly in Weyl semi-metals: Application to  Pyrochlore Iridates}

\author{Vivek Aji}
\address{Department of Physics and Astronomy, University of California,
Riverside, CA 92521}
\begin{abstract}
Weyl semimetals are three dimensional analogs of graphene where the energy of the excitations are a linear function of their momentum. Pyrochlore Iridates $(A_{2}Ir_{2}O_{7})$ with A =yttrium or lanthanide element) are conjectured to be examples of such a system, with the low energy physics described by twenty four Weyl nodes.  An intriguing possibility is that these materials provide a physical realization of the Adler-Bell-Jackiw anomaly. In this letter we investigate the properties of pyrochlore iridates in an applied magnetic field. We find that the dispersion of the lowest landau level depends on the direction of the applied magnetic field. Consequently the velocity at low energies can be manipulated by changing the \emph{direction} of the applied field. The resulting anisotropy in longitudinal conductivity is investigated.  
\end{abstract}
\maketitle

Graphene\cite{geim} and topological insulators\cite{kane,fu1,roy} have provided the venue for condensed matter realizations, outside of liquid Helium\cite{volovik},  of nontrivial phenomena originally in the realm of high energy physics. Massless relativisitic fermions\cite{ahcn}, Klein tunneling \cite{ahcn}, Theta vacuum (i.e. axion electrodynamics) \cite{hasan}, and Majorana modes \cite{hasan, fu} are a few examples. A common feature of these systems is that the low energy physics is described by a two component Dirac Hamiltonian with the fermionic momentum is confined to two dimensions. Recently pyrochlore iridates have been conjectured to realize the Adler-Bell-Jackiw (ABJ)\cite{adler, bell} chiral anomaly, adding another example to the growing list.

\begin{figure}
\includegraphics[height=0.6\columnwidth, clip]{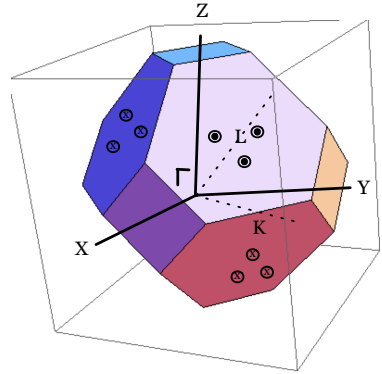}
\caption{The low energy physics of pyrochlore iridates is described by linearly dispersing fermionic modes near nodes in the band structure. The position of Weyl nodes in the Brillouin zone is shown in the figure. There are three nodes, with the same chirality, located in the vicinity of the $L$ points, and  nodes related by inversion have opposite chirality.}
 \label{nodes}
\end{figure}

Wan et al. [\onlinecite{ashvin_weyl}] explored the possibility of the three dimensional analog of graphene being realized in the pyrochlore iridates. These materials have a large spin orbit couplings and are in a regime of intermediate correlations, making them promising candidates to realize topological insulators\cite{shitade, pesin}. They have a magnetic ground state\cite{nobuyuki,kazuyuki} and, within a LSDA+U+SO calculations, conjectured to be semi-metals. Most strikingly the low energy physics is described by the Weyl equation, which is the two component version of the Dirac equation. There are 24 Weyl nodes, three around each $L$ point ($\left[111\right]$ and equivalent directions) in the Brilloiun zone (see fig.\ref{nodes}). Nodes related either by inversion or reflection about the $\{xy,yz,zx\}$ planes have opposite chirality. Consequently the material is  expected to have an anomalous hall response to applied uniaxial pressure and is susceptible to charge ordering in large magnetic fields\cite{yran}.

In a quantizing magnetic field, the Lowest Landua Level (LLL) is a linear function of the magnitude of a momentum, with the sign determined by the band structure. If an electric field is applied parallel to the magnetic field, the Adler-Bell-Jackiw (ABJ)\cite{adler, bell} axial anomaly leads to an anomalous magneto-conductance. The origin of this effect is in the production of  Weyl fermions of a given chirality and an equivalent annihilation of the opposite chirality\cite{nielsen}. This translates to a transfer of particles from one Weyl node to another of opposite chirality at a constant rate. To reach a steady state, this is balanced by inter-node scattering due to impurities.

In the absence of the magnetic field, the intra-node scattering is quite effective in relaxing the momentum. In the presence of a large field, such that only the LLL is occupied, intra-node scattering is suppressed due to a lack of phase space. The only scattering mechanism available are processes that involve different nodes of opposite chirality. Since these nodes are located at different point in the Brillouin zone, Nielsen and Ninomeya \cite{nielsen}  argued that  the corresponding scattering rate is much smaller implying a large magneto-conductivity. 

In this letter we focus on the anomalous magneto-conductivity expected in Weyl semi-metals \cite{nielsen, ashvin_weyl}. In general the Hamiltonian takes the form $
\pm\vec{q}\cdot\textbf{V}\cdot\vec{\sigma}$ where $\vec{q}$ is the momentum, $\textbf{V}$ is a real matrix,  $\pm$ label right handed (RH) and left handed (LH) chirality,  and $\vec{\sigma}$ $=\{\sigma_{x}, $ $\sigma_{y},$ $\sigma_{z}\} $. We first address the question of the dispersion of the LLL for arbitrary $\textbf{V}$. We find that the energy is determined by the component of the momentum parallel to the applied field. Rather surprisingly the velocity of the mode depends on the direction of the magnetic field. This implies that the low energy dispersion and density of states can be manipulated by varying the direction of the magnetic field.

The low energy Hamiltonian in the vicinity the Weyl nodes of the proposed topological semimetal state is 

\begin{eqnarray}\label{ham}
H\left(\vec{q}\right) &=& \left(\Delta + {q_{z}^{2}\over{2m_{1}}}-{q_{\perp}^{2}\over{2m_{2}}}\right)\sigma_{z} \\ \nonumber &+&\left(\beta q_{z}+c_{1}q_{\perp}^{3}\cos\left(3\theta\right)\right)\sigma_{y} + c_{2}q_{\perp}^{3}\sin\left(3\theta\right)\sigma_{x}
\end{eqnarray}
where the local $z-$axis is taken along the $\Gamma-L$ direction and the local $x-$axis along the $L-K$ direction in the Brillouin zone. Notice that these direction rotate with respect to the global coordinates from one $L$ point to another. At the Weyl nodes, there is a degeneracy among two states with opposite symmetry under inversion. The physics of the two state system is captured by the $\sigma$ matrices.  Define ${q}_{0z}$ and $\vec{q}_{0\perp}$ as the displacements of the node away from the $L$ point parallel and perpendicular to the $\Gamma-L$ line respectively. $\theta$ is defined as the angle $\vec{q}_{0\perp}$ makes with the $x-$ axis. The nodes are located at $\theta = p\pi/3$, $q_{0z} = \pm c_{1}q_{0\perp}^{3}/\beta$ (positive(negative) for odd (even) values of $p$ ), and $q_{0\perp}$ satisfying the equation $\Delta$ $+ q_{0z}^{2}/{2m_{1}}$ $-q_{0\perp}^{2}/{2m_{2}}$ $=0$. 

In the vicinity of these nodes the Hamiltonian can be expanded and written as 
\begin{equation}
H_{i}\left(\delta \vec{q}\right) \approx \delta\vec{q}\cdot \textbf{V}_{i}\cdot\vec{\sigma} 
\end{equation}
where $\delta\vec{q} = \vec{q}-\vec{q}_{0}$, and $\textbf{V}_{i}$ is a real matrix whose entries depend on the node index $i$. For the Hamiltonian in eq.\ref{ham} the matrix $\textbf{V}$ is generically not symmetric and does not have all real eigenvalues.  This is quite unlike the case considered in the context of the ABJ anomaly\cite{nielsen} or the  anomalous hall effect \cite{yran}. For a diagonal $\textbf{V}$ matrix, the direction of the momentum, whose magnitude determines the energy if the lowest Landau level, is parallel to the applied magnetic field. Since momentum does not commute with the electromagnetic vector potential, and the fact that the latter is always transverse to the $\vec{B}$ field, suggest that this property is generally valid. To understand the nature of the ABJ anomaly in iridates, we first verify that this conjecture holds for arbitrary $\textbf{V}$ matrices. 

The procedure for computing the energy is sketched out here (details provided elsewhere \cite{aji_weyl}). Consider a single Weyl node in an applied magnetic field, $\vec{B}$. We can always rotate the coordinate system so that the $\vec{B}$ lies in the $xz-$plane. In this reference frame $\vec{B} = B\left\{\sin\left(\theta\right),0,\cos\left(\theta\right)\right\}$. Using a Landau gauge we write the corresponding vector potential as $B\left\{-\cos\left(\theta\right)y,0,\sin\left(\theta\right)y\right\}$. Given this choice, the system is translationally invariant in the $x$ and $z$ directions. The  wave-function of the LLL has the form $\left\{u, v\right\}\phi\left(y\right)e^{-i \delta k_{x}x - i \delta k_{z} z}$, where $u$ and $v$ are constants. Rather remarkable, the energy $\epsilon_{0}$, calculated with this choice of gauge, can be written in a gauge invariant form as

\begin{eqnarray}\label{lll}
\epsilon_{0} &=& -{Det\left[\textbf{V}\right] \over{\|adj\left[\textbf{V}\right]\cdot\vec{B} \|}}\delta\vec{q}\cdot \vec{B}\\ \nonumber
&=& -sgn(Det[\textbf{V}]){\delta\vec{q}\cdot \vec{B}\over{\|\textbf{V}^{-1}\cdot\vec{B} \|}}
\end{eqnarray}
where $Det\left[\textbf{V}\right] $ and $adj\left[\textbf{V}\right]$ are the determinant, and adjugate of the matrix $\left[\textbf{V}\right] $. The sign of the dispersion, and hence the chirality, is determined by the determinant. The energy is inversely proportional to the projection of the deviation of the momentum in the direction of the applied field. 

The results in eqn.\ref{lll} is the key result and we will explore the consequences in the rest of the letter. The most important feature of the dispersion is the dependence of the velocity of the low energy excitations on the direction of the applied field. This anisotropy is inherited from the underlying band structure. While most systems have inherent anisotropies, what makes Weyl semi-metals unique is that they have linearly dispersing modes, the left handed and right handed branches of which are located at distinct point in the Brillouin zone. All small momentum scattering are suppressed leading to a large magneto-conductivity. Furthermore the ability to change the velocity implies that the low energy density of states can also be manipulated. As such all thermodynamic and transport properties are sensitive to the direction of the applied field. Here we focus on the ABJ anomaly.

Let us now consider the case of the pyrochlore iridates. Given the Hamiltonian (eqn.\ref{ham}) we construct the relevant matrices. To make further progress we use parameters that best fit the LSDA+U+SO calculations \cite{ashvin_weyl, yran}: $m_{1}$= $m_{2}$ $=0.5$ eV$^{-1}$, $c_{1}$ $=c_{2}$ $=1.0$ eV, $\beta = 0.5$ eV, $\Delta$ $=0.18$ eV ($\vec{q}$ is dimensionless). These parameters give $q_{0\perp} = 0.48$ and $q_{0z}= \pm 0.22$. 

To get insight on the general properties let us look at the nodes near  $\left[111\right]$. The matrices for  $\theta =\{0,2 \pi/3, -2\pi/3\}$ are

\begin{eqnarray}\label{v111}
\textbf{V}_{\theta=0}&=&\left(\begin{array}{ccc}0 & 3 q_{0\perp}^{2} &- 2 q_{0\perp} \\3 q_{0\perp}^{2} & 0 & 0 \\0 & {1\over 2} & 2 q_{0z}\end{array}\right)\\ \nonumber
\textbf{V}_{\theta=\pm 2\pi/3}&=& \left(\begin{array}{ccc}\mp{3\sqrt{3}\over 2} q_{0\perp}^{2} & -{3\over 2} q_{0\perp}^{2} & q_{0\perp}  \\-{3\over 2} q_{0\perp}^{2} & \pm{3\sqrt{3}\over 2} q_{0\perp}^{2} & \mp \sqrt{3}q_{0\perp} \\0 & {1\over 2} & 2 q_{0z}\end{array}\right)
\end{eqnarray}
The system possesses three fold rotation symmetry about the $\Gamma-L$ axis and the three matrices are related by 120 degree rotation about the $z-$axis. If $\textbf{U}$ is a rotation matrix for $2\pi/3$ rotation about the local $z-$axis, than $\textbf{U}\textbf{V}_{\theta=0}$ $ = \textbf{V}_{\theta=2\pi/3}$, $\textbf{U}\textbf{V}_{\theta=2\pi/3}$ $ = \textbf{V}_{\theta=-2\pi/3}$ and $\textbf{U}\textbf{V}_{\theta=-2\pi/3}$ $ = \textbf{V}_{\theta=0}$. In other words, knowing one is sufficient to generate the others. The determinant of these three matrices are all equal and given by $Det[\textbf{V}]_{i}$ $ = -3 q_{0\perp}^{3}$ $-18 q_{0\perp}^{4} q_{0z}$.

Since $\textbf{V}^{-1} =adj\left[\textbf{V}\right]/Det\left[\textbf{V}\right]$, the adjugate matrices have the same transformation properties as the inverse matrix. Given the three matrices near the $[111]$ point, all other can be constructed by symmetry. The matrices for the Weyl points related by inversion to those near $[111]$ are obtained by changing the sign of the third column or equivalently the sign of $\sigma_{z}$. This remains true for all nodes related by inversion as the local $z-$axis changes sign.  The nodes near $[\bar{1}\bar{1}1]$ have the same structure as those near $[111]$ while those near $[\bar{1}11]$ have the sign of the first column changed. This ensures the geometry and helicity obtained within  LSDA+U+SO calculations \cite{ashvin_weyl}.
  
Pyrochlore iridates have cubic symmetry. Thus the anomalous Hall is zero unless an uniaxial pressure is applied to break the symmetry\cite{yran}. The same property leads to an isotropic density of states as a function of the direction of the magnetic field. To get anomalous response in thermodynamic properties from lowest Landau level, one needs to have Weyl nodes in systems that are inherently anisotropic. The layered heterostructure of normal and topological insulators\cite{burkov} is one example of such systems. Remarkably, even for an isotropic system the transport properties can be anisotropic as we show below.  
  
Since the dispersion of the LLL is in the direction of the applied magnetic field, only  the response to an electric field, $\vec{E} = E_{0}\hat{B}$ applied parallel to the magnetic field will be considered. We will assume that the magnetic field is strong enough so that only the LLL is occupied. The dispersion being linear, elastic scattering within a single node is suppressed due to the lack of phase space. Stated differently no momentum relaxation is possible within a node as states with opposite velocities do not exist.  However the scattering between two nodes cannot be ignored. In the presence of the electric field, there is a generation of RH particles and annihilation of LH particles. The rate of production is given by the rate of change of energy, $v\delta\dot{q}$,  times the density of states, $eB/v\hbar$.  Since $\delta\dot{q} = eE_{0}$, we get $\dot{{N}}_{||} $ $=\delta \dot{q}eB/\hbar$ $= e^{2} E_{0} B/\hbar$. This rate has to be balanced by the scattering between nodes with the opposite chirality to maintain a steady state.  

Weyl nodes always come in pairs with opposite chirality. Before considering all 24 Weyl nodes, we first look at a single pair whose energies are given by $\epsilon_{0} = \pm v \delta q_{||}$. For impurity scatterers, where the transition probability between states with different moment is independent of their momenta, the scattering rate, $\tau^{-1}$ is proportional to the density of states and is given by

\begin{equation}
{1\over \tau} = C{eB\over {v\hbar}}
\end{equation}
where $C$ is a constant determined by the strength of scattering potential. We will comment on more general forms of scattering later. The ABJ anomaly requires an imbalance in particle number at the RH and LH Weyl nodes. If the difference in chemical potential between the RH and LH nodes is $\Delta \mu$, than energy balance requires

\begin{equation}\label{mu}
\Delta\mu  = eE_{0} v\tau
\end{equation}
This is because over the scattering time $\tau$, the momentum changes by $eE_{0}\tau$. Since the change in energy is $v$ times change in momentum, the net energy transferred from one cone to the other is $eE_{0}\tau v$. A steady state is achieved if this transfer is balanced by the difference in chemical potential. 

The difference in chemical potential leads to a current given by 

\begin{eqnarray}\label{abj}
J_{A} &=& nev\\ \nonumber
J_{A} &=& {eB\over {v\hbar}}\Delta\mu e v\\ \nonumber
&=& e (eB/\hbar)eE_{0}v\tau\\ \nonumber
&=&C^{-1}e^{2}v^{2}E_{0}
\end{eqnarray}
The subscript $A$ refers to the anomalous response. So far we have only considered momentum independent scattering rate. Let us look at the response in the presence of screened charged impurities characterized by matrix elements of the form $1/({q^{2}+\kappa^{2}})$. The internode scattering probes intermediate to large momenta, as the nodes are physically separated in the Brillouin zone. The transition probabilities, which are proportional to the square of the matrix elements,  fall of as $1/|\vec{q}|^{4}$ giving a large $\tau$. In contrast, the conductivity in zero magnetic field is much smaller as it is dominated by intra-node scattering ($\sim 1/\kappa^{4}$). Thus the conductivity in the presence of the magnetic field can be much larger than that in zero field. The results in eq.\ref{mu} and eq.\ref{abj}, and the argument of large magneto-conductance, are the main conclusions of Nielsen and Ninomeya [\onlinecite{nielsen}].

Having reviewed the expected nature of the magneto-conductance, we return to the discussion of the iridates. The key feature of eq.\ref{abj} is its dependence on the velocity $v$. As we have seen in our discussion of the dispersion of the LLL, the velocity can be tuned by changing the direction of the applied field. Moreover the different Weyl nodes have different Fermi wave-vector and different density of states.  Since the longitudinal conductivity depends on the scattering rate, we study two cases. We focus on momentum independent processes which either scatter only between nodes with opposite velocities or only between nodes of opposite chirality. Detailed studies of how various scattering processes will reveal the underlying anisotropy will be part of future efforts, but all of them are sensitive to the underlying dispersion which is the source of the phenomena.

\begin{figure}
 \includegraphics[width=0.75\columnwidth, clip]{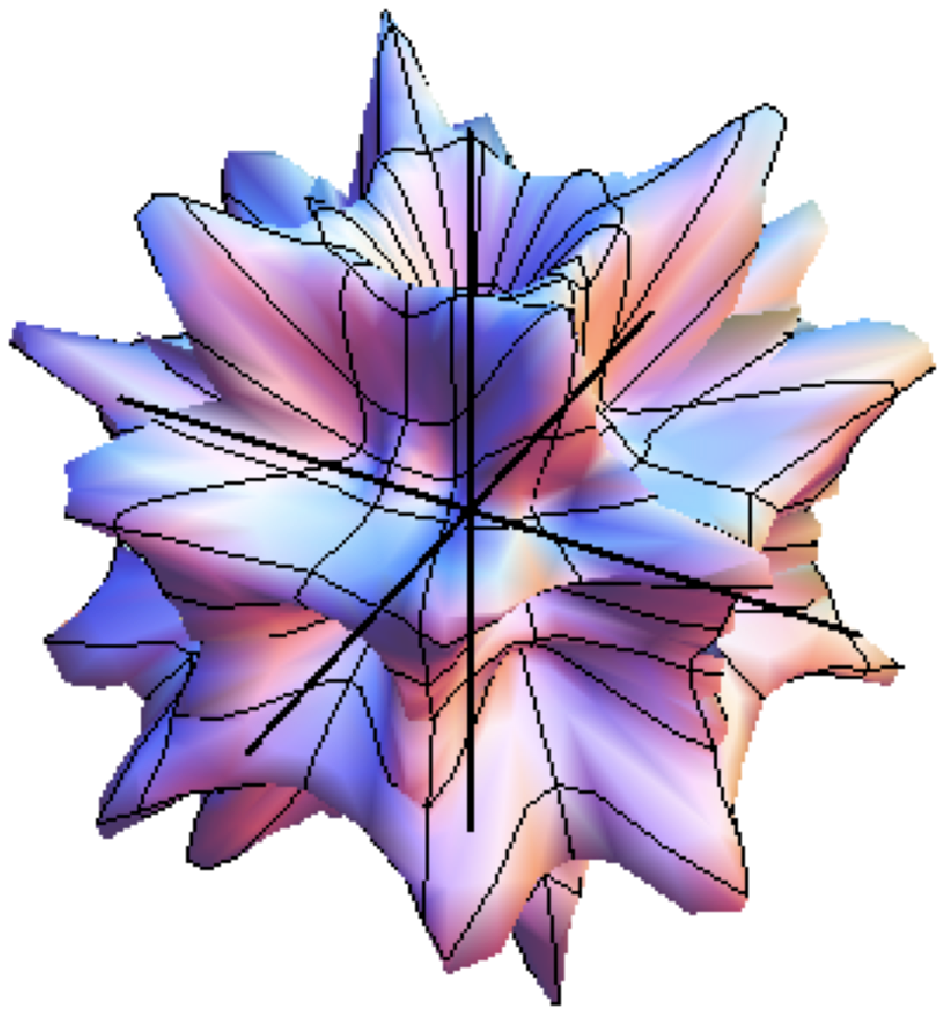}
\includegraphics[width=0.75\columnwidth, clip]{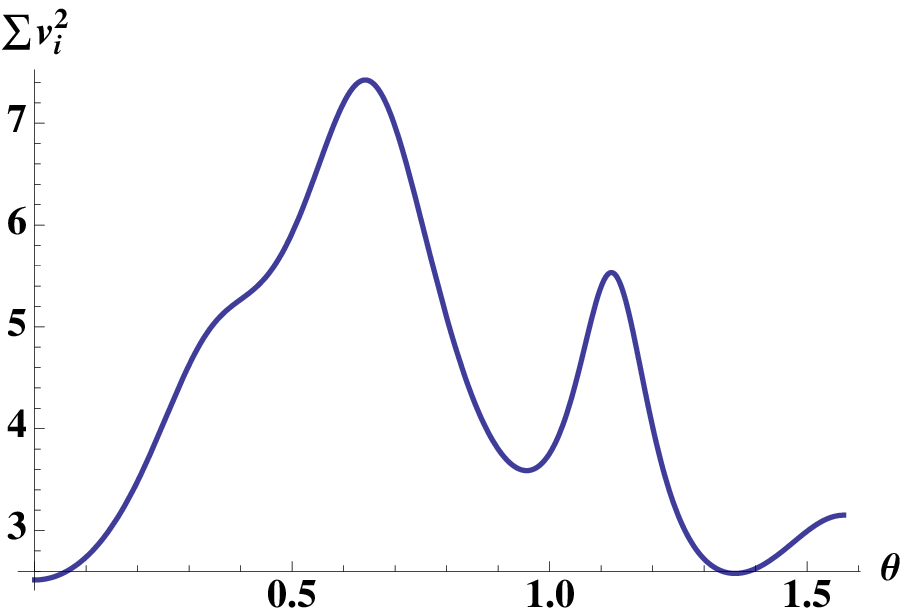}
\caption{Top: The variation of the conductivity, for scattering potential that only couple nodes that are related by inversion, as a function of the direction of magnetic field is shown as a three dimensional spherical plot. The radial distance of a point on the figure from the origin is a measure of the conductivity for the magnetic field along that direction. Bottom:  A cut along the $Z-K$ direction is shown.}
 \label{cond1}
\end{figure}

Weyl nodes related by time reversal have opposite velocities and the bands are related by inversion. For scattering processes that conserve the corresponding quantum numbers, scattering between these nodes dominate.  The scattering rate is proportional to the density of states of the nodes and the total conductivity is $\sigma = \sum_{i}C^{-1}e^{2}v_{i}^{2}$. In fig.\ref{cond1} the conductivity is plotted as a function of the magnetic field and a cut along the $Z-K$ direction, where both the maximum and minimum values are obtained, is shown. The anisotropy, define as $(\sigma_{max}-\sigma_{min})/(\sigma_{max}+\sigma_{min})$ is $\sim 50\%$.

\begin{figure}
 \includegraphics[width=0.75\columnwidth, clip]{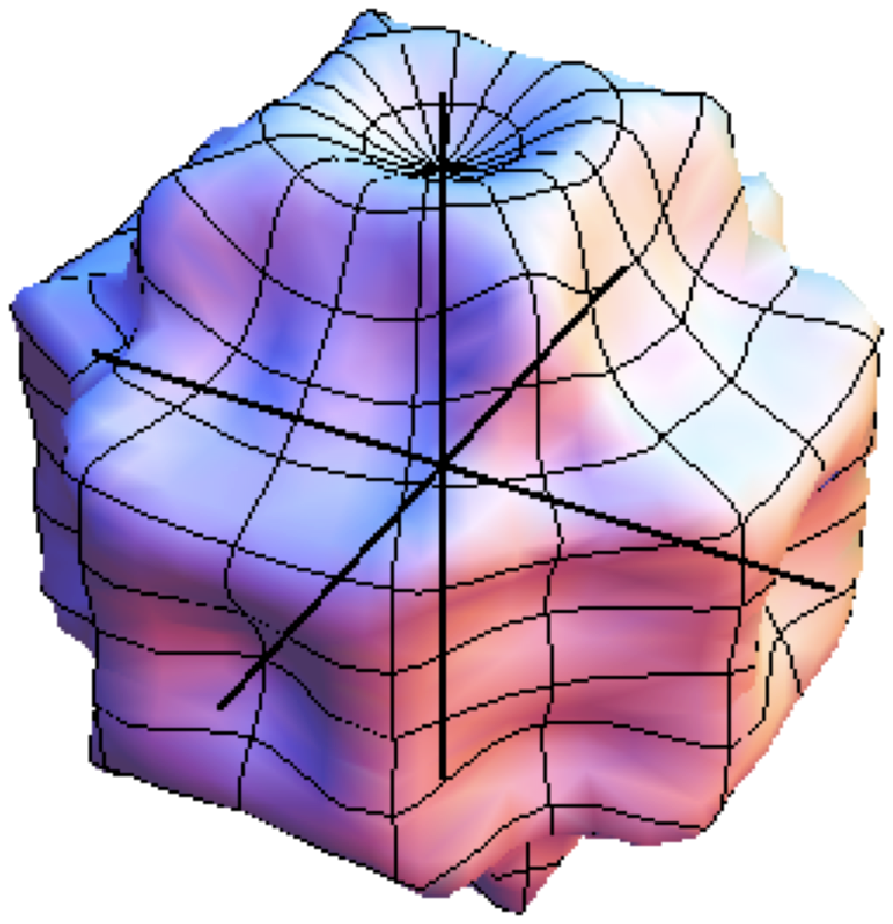}
\includegraphics[width=0.75\columnwidth, clip]{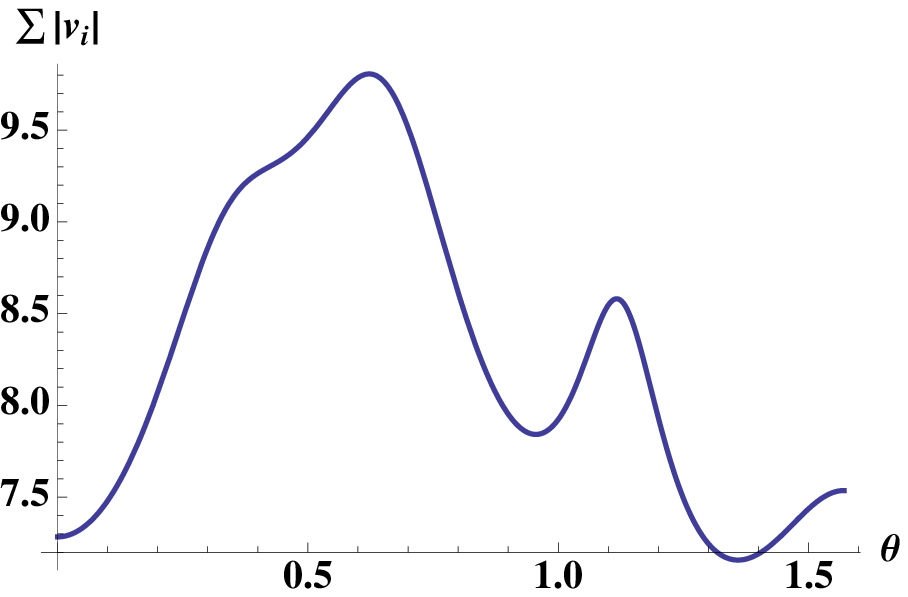}
\caption{Top: The variation of the conductivity, for scattering potentials that couple all nodes, as a function of the direction of magnetic field is shown as a three dimensional spherical plot. The radial distance of a point on the figure from the origin is a measure of the conductivity for the magnetic field along that direction. Bottom: A cut along the $Z-K$ direction is shown.}
 \label{cond2}
\end{figure}

For scattering processes that do not preserve any of the lattice symmetries, all nodes participate in momentum relaxation of a given node. In this case the scattering rate will be proportional to one half of the total density of states. The factor of half accounting for they fact that only half of the nodes have the opposite velocity, and can contribute to momentum relaxation. In fig.\ref{cond2} the results for the anisotropy of conductivity is presented for this case. The anisotropy is $\sim 17\%$.

We note that the direction dependence in transport is obtained even in a system that has cubic symmetry. For topological-normal insulator heterostructure where $[\textbf{V}]$ is diagonal, but has different velocities in-plane as opposed to perpendicular to the plane, the density of states is also anisotropic. In such systems even the low energy thermodynamics will depend on the direction of the applied field. A more general study of the low energy behavior of specific heat, susceptibility and the effect of momentum dependent scattering will be part of a future effort in characterizing the novel behavior of Weyl semi-metals due to their unique zero Landau level dispersion.  

In this letter we have considered the nature of magneto-conducatnce in Weyl semi-metals. In the presence of an applied field the energy of the LLL is obtained. We find that the the level always disperses linearly with respect to the momenta parallel to the applied field. This result is general and holds even for systems with arbitrary Weyl hamiltonians. The velocity of the low energy modes depends on the direction of the applied field. This feature is exploited in the context of the pyrochlore iridates. Using the symmetries of the crystal, we derive the Weyl equation at the 24 nodes. Applying magnetic fields in different directions allows us to manipulate the low energy physics, leading to anisotropic transport properties.

VA acknowledges the support of University of California at Riverside through the initial complement. V.A. thanks A.Vishwanath for helpful suggestions and comments.


\end{document}